\documentclass[twocolumn,10pt]{article}
\usepackage{url}
\usepackage{latex8}
\usepackage{times}
\usepackage{ifpdf}
\usepackage{xspace}

\ifpdf
\usepackage[pdftex]{graphicx}
\else
\usepackage{graphicx}
\fi

\ifpdf
\DeclareGraphicsExtensions{.pdf, .jpg, .tif}
\else
\DeclareGraphicsExtensions{.eps, .jpg}
\fi

\usepackage{latexsym}

\usepackage[symbol]{footmisc}
\DefineFNsymbols{irene}[text]{{$\ast$} {$\star$} {$\circ$} {$\bullet$}{$I$} {II} {III} {IV} {V}}
\setfnsymbol{irene}

\newcommand{\comment}[1]{{}}

\newcommand{\conference}{SEA\xspace}

\begin{document}

\title{{\Large{\bf Experimental Evaluation of Algorithms}\\ for the Food-Selection Problem}}

\author{\addtocounter{footnote}{0}
{\em Camil Demetrescu}~\thanks{Department of Computer, Control, and Management Engineering,
Sapienza University of Rome, Via Ariosto
XXV (25), CLXXXV (00185) Roma, Italy.  E-mail:
{\tt demetres}{\tt @dis.uniroma1.it}.}~\\ \and 
\addtocounter{footnote}{+0}
{\em Irene
Finocchi}~\thanks{Department of Computer Science,
Sapienza University of Rome, Via Salaria
CXIII (113), CXCVIII (00198) Roma, Italy.  E-mail:
{\tt finocchi@di.uniroma1.it}.}~\\
\and {\em Giuseppe F. Italiano}~\thanks{Department Civil Engineering and Computer Science Engineering,
University of Rome ``Tor Vergata,'' Via del Politecnico
I, CXXXIII (00133) Roma, Italy.  E-mail:
{\tt italiano@disp.uniroma2.it}.}
\and {\em Luigi Laura}~\thanks{Most of the time his car, traveling between the Department of Computer, Control, and Management Engineering,
Sapienza University of Rome and the Department of Civil Engineering and Computer Science Engineering,
University of Rome ``Tor Vergata.''  E-mail:
{\tt laura@dis.uniroma1.it}.}
}

\date{}

\maketitle

\begin{abstract}
\noindent In this paper, we describe the result of our experiments on
Algorithms for the Food-Selection Problem, which is the fundamental
problem first stated and addressed in the seminal paper~\cite{pigout}.
Because the key aspect of any experimental evaluation is the
\textbf{reproducibility}, we detail deeply the setup of all our
experiments, thus leaving to the interested eater the opportunity to
reproduce all the results described in this paper. More specifically, we
describe all the answers we provided to the
questions proposed in~\cite{pigout}:

\vspace{-2.4mm}

\begin{itemize}

\item Where can I have dinner tonight?

\vspace{-2.4mm}

\item What is the typical Roman cuisine that I should (not) miss?

\vspace{-2.4mm}

\item Where can I find the best coffee or gelato in town? 

\end{itemize}
\vspace{-1mm}


\bigskip


\end{abstract}

\vspace{-3mm}

\section{Introduction}
\label{se:intro}

\vspace{-2mm}
Eating is one of life's greatest pleasures, especially in Rome. Thus, 
the general advice recommends foreigners to be adventurous and to 
follow the well-known maxim:
\begin{center}
{\em When in Rome, do as the Romans do.}
\end{center}
Unfortunately, previous experimental evidence suggests that this 
approach may not even lead to a feasible solution in finite time because
of convergence problems.
Indeed, there is an incredible variety of eateries, especially in 
districts such as {\em Testaccio}, {\em Trastevere}, {\em Campo de' 
Fiori}, {\em Piazza Navona}, or the {\em Ghetto}. The problem of 
selecting a good restaurant seems therefore of practical importance 
in this mouth-watering scenario.

In this paper we present the experimental evaluation of the algorithms for the food selection 
problem originally presented in \cite{pigout}. These algorithms are based on fundamental algorithmic 
techniques including random sampling, divide et impera (sorry guys, 
but we are in  Rome!), local search, hill climbing, and the greedy 
approach, and require only constant time. However, given that every month 
is high season for tourism in Rome, if you don't plan in advance you 
may be forced to use backtracking (the restaurant may be full or 
closed): this occurrence could increase considerably your running 
time.

\begin{figure*}[t!]
\begin{center}
\includegraphics[width=0.85\textwidth]{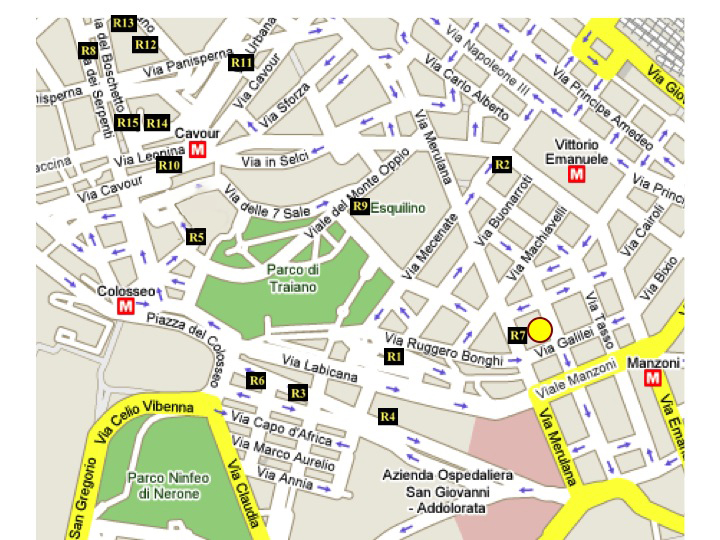}
\end{center}
\caption{Food selection by local search. The yellow point is the
\conference conference venue.\label{fi:localsearch}}
\end{figure*}

\section{Local Search}
\label{se:localsearch}

Well, it's VIII PM and \conference sessions are over. You feel so hungry that 
even power-law graphs look like fettuccine. Local search seems the 
appropriate technique to apply in this case. Like any good Roman 
district, you will not be spared by the wealth of great restaurants, 
trattorias and pizzerias around the \conference venue. We enumerate now all the solutions we found using a simple local search algorithm. 
All the solutions can be easily reached on foot (in XV minutes or so, 
depending on hunger pangs) even if you don't like walking or if you 
are not very good at computing shortest paths: see 
Figure~\ref{fi:localsearch}.  

\begin{enumerate}

\item[R1] {\em Il Patio Pizzeria}, Via Labicana 29, ph: 06-7003406. Here you should definitely try their gnocchi with cheese!

\item[R2] {\em Vecchia Roma (Old Rome)}, Via Ferruccio 12c, ph: 
06-4467143, closed on Sunday. Don't miss the bucatini all'amatriciana served in a hollow pecorino cheese.

\item[R3] {\em Naumachia},  V. Celimontana 7, ph: 
06-7002764. A must is their hand-made pasta with a soft sausage sauce. 

\item[R4] {\em Hostaria Isidoro}, San Giovanni in Laterano 59, ph: 
06-7008266. If you like pasta, this place is a good choice for you!

\item[R5] {\em Hostaria da Nerone}, Via delle Terme di Tito 96, ph: 06-4745207, closed on Sundays. This is a tiny, friendly restaurant.  

\item[R6] {\em La Pace del Cervello (The Rest of the Brain)}, Via Dei Ss. Quattro 63, ph: 06-7005173. This is a (good!) pizzeria and a pub; it is open up to late night.

\item[R7] {\em Beer House}, Via Merulana 109, ph: 
06-7096265. It is a pub with home made roman cuisine! Very close to conference venue!

\item[R8] {\em Il Grillo Brillo (Drunken Cricket)}, Via dei serpenti 79, ph: 06-4883029. Good meat and fish.

\item[R9] {\em Kick-Off}, Via Delle terme di Traiano 4, ph: 06-48904343. This is a pizzeria, a pub, and it has also two regular five-a-side soccer fields, just in case you want to play soccer before eating!

\item[R10] {\em Pizzeria Borgia}, Via Cavour 215, ph: 06-4743538. Good pizzeria, with an inside garden.

\item[R11] {\em Robin Hood}, Via Cavour 162, ph: 
06-48903511. Here you can try their famous ``Pizza Show'': a complete dinner all served on top of pizzas!

\item[R12] {\em Carbonara}, Via Panisperna 214, ph: 06-4825176, closed
on Sundays. Traditional roman dishes, nice atmosphere, in the famous
street that gave its name to the \emph{Via Panisperna boys}, a group of
young scientists led by Enrico Fermi; here in 1934 they discovered the
``slow neutrons.''

\item[R7] {\em Al Giubileo}, Via del boschetto 44, ph: 
06-4818879. Here they serve ``pizza no stop'': all you can eat pizza!


\item[R14] {\em Pizzeria Chicco Di Grano}, Degli Zingari 6, ph: 06-47825033. Tuscany style restaurant in the middle of Rome!

\item[R15] {\em Al Vino al vino}, Wine Bar, Via dei Serpenti 19, ph: 06-06485803. More than 600 different wines! 

\end{enumerate}

%








\section{Hill Climbing}
\label{se:hill-climbing}

Hill climbing appears to be a very promising approach: to understand 
why, let's review a little bit of history on the VII Hills of Rome.
Tradition says that Rome was founded in DCCLIII (753) BC by twin 
boys, Romulus and Remus. Romulus and Remus were abandoned by their 
parents, put in a cradle and sent off to be drowned on the Tiber 
river. The boys were found by a she-wolf who took care of them as if 
they were her own cubs. (We remark that all of this happened well 
before Mowgli appeared.) When the boys grew a bit older, a shepherd 
found them: he and his wife raised them to be young men. At that 
point, the twins set out on a quest: on their way they found a 
beautiful land with seven hills, which they named Palatine 
(Palatino), Aventine (Aventino), Capitoline (Campidoglio), Quirinal 
(Quirinale), Viminal (Viminale), Esquiline (Esquilino) and Caelian 
(Celio). They were very excited: but, who would rule that piece of 
land? To determine this, Romulus stood on the Aventine hill 
and Remus stood on the Palatine: whichever hill the birds flew over, 
that brother would rule the land. It was said to be a message from 
the gods. After a long time, six birds flew over the Palatine hill 
where Remus was standing, and so he thought he would rule, until 
twelve birds flew over the Aventine hill where Romulus stood, and 
Romulus ended up ruling Rome. 

Nowadays, the seven hills, the valleys in between, and some other 
hills such as Janiculum (Gianicolo, rising above Trastevere) and 
Pincian (Pincio, rising above Piazza del Popolo) represent just the 
central part of the city (Centro Storico), in which you will find the 
main monuments, archeology sites, wonderful views, museums, and fine 
restaurants.  But don't be scared about climbing Rome's hills: they 
aren't that big after all! Some restaurants and trattorias in central 
Rome are listed below.

\begin{enumerate}

\item[R16] {\em Antico arco}, Gianicolo, Piazzale Aurelio 7, ph: 
06-5815274, closed on Sunday (A2)

\item[R17] {\em Arancia blu}, San Lorenzo, via dei Latini 55-65, ph: 
06-4454105, vegetarian  (east of C2)

\item[R18] {\em Da Armando al Pantheon}, Centro Storico -- Pantheon, 
salita de' Crescenzi 31, ph: 06-68803034, closed on Sunday (B2)

\item[R19] {\em Da Benito e Gilberto}, Vaticano, via  del Falco 19, 
ph: 06-6867769, closed on Sunday/Monday, very good fish (A1)

\item[R20] {\em Caf\`e Mancini}, Centro Storico -- Campo Marzio, via  
Metastasio 21, ph: 06-6872051, closed on Sunday, once in the centre, 
reach it by walking along some of the most characteristics streets in 
Rome (B2)

\item[R21] {\em La Campana}, Centro Storico -- Campo Marzio, vicolo 
della Campana 18, ph: 06-6867820, closed on Monday, some people say 
this is the most ancient restaurant in Rome... (B2)

\item[R22] {\em Camponeschi}, Centro Storico -- Campo De' Fiori, 
Piazza Farnese 50/50a, ph: 06-6874927, closed on Sunday, historical 
Roman restaurant in a wonderful location, expensive (A2)

\item[R23] {\em Checchino dal 1887}, Testaccio, Via di Monte 
Testaccio 30, ph: 06-5743816, closed on Sunday/Monday, another 
historical Roman restaurant (A3)

\item[R24] {\em Checco Er Carrettiere}, Trastevere, Via Benedetta 10, 
ph: 06-5800985, closed on Sunday (A3)

\item[R25] {\em Il Convivio}, Centro Storico -- Navona, Vicolo dei 
Soldati 31, ph: 06-6869432, closed on Sunday, very good restaurant, 
expensive (A2)

\item[R26] {\em Dar Cordaro}, Trastevere, Piazzale Portuense 4, ph: 
06-5836751, closed on Sunday/Monday (A3)

\item[R27] {\em Il Drappo}, Centro Storico -- Campo De' Fiori, Vicolo 
del Malpasso 9, ph: 06-6877365, closed on Sunday, Sardinian cuisine 
(A2)

\item[R28] {\em Gaud\`i}, Parioli, Via R. Giovannelli 8, ph: 
06-8845451, pizza (north of C1)

\item[R29] {\em Da Giggetto al Portico d'Ottavia}, Antico Ghetto, Via 
Del Portico d' Ottavia 21/a	, ph: 06-6861105, closed on Monday (B2)

\item[R30] {\em Hostaria degli Artisti}, Esquilino, Via G. Sommeiller 
6/8, ph: 06-7018148, Neapolitan cuisine (C2)

\item[R31] {\em Matricianella}, Centro Storico -- Campo Marzio, via  
del Leone 2/4, ph: 06-6832100, closed on Sunday, authentic Roman 
cuisine at convenient prices (B2)

\item[R32] {\em La Penna d'Oca}, Centro Storico -- Popolo, via  della 
Penna 53, ph: 06-3202898, closed on Sunday, for hard to please 
gourmets (B1)

\item[R33] {\em Dar Poeta}, Trastevere, Vicolo del  Bologna 45, ph: 
06-5880516, very good pizza (A3)

\item[R34] {\em Sora Lella}, Centro Storico -- Isola Tiberina, Via di 
Ponte Quattro Capi 16, ph: 06-6861601, famous restaurant in a 
position unique in the world (B2)

\item[R35] {\em Tram Tram}, San Lorenzo, Via dei Reti 44, ph: 
06-490416 (east of C2)

\item[R36] {\em Il Tulipano Nero}, Trastevere, Via Roma Libera 15,
ph: 06-5818309, good pasta and pizza (A3)

\end{enumerate}

\begin{figure}[h!]
\begin{center}
\includegraphics[width=0.45\textwidth]{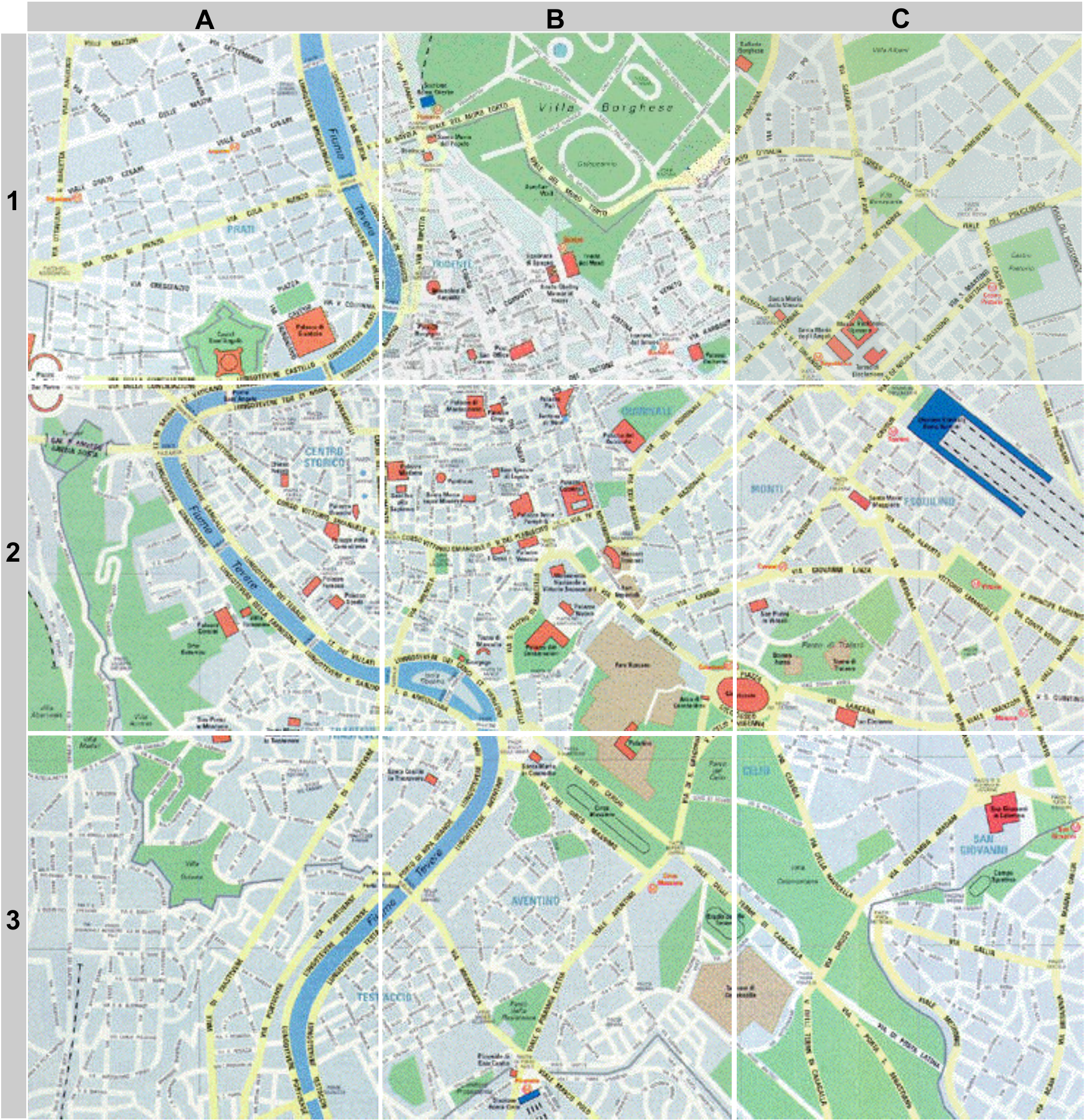}
\end{center}

\caption{Geometrical decomposition used by our divide et impera food 
selection algorithm (see Section~\ref{se:divideetimpera}).}
\label{fi:mappa-large}
\end{figure}

\begin{figure*}[t]
\centerline{
\begin{tabular}{|c|l|} \hline 
A1 & Prati - San Pietro  - Castel Sant'Angelo - Vaticano \\ \hline
A2 & Piazza Navona - Piazza Farnese - Campo De' Fiori \\
   & Via Giulia - Gianicolo \\ \hline
A3 & Trastevere - Testaccio \\ \hline
B1 & Piazza di Spagna - Via Veneto  - Via Condotti - Tritone \\ \hline
B2 & Quirinale - Fontana di Trevi - Fori Imperiali - Mercati Traianei 
\\
   & Foro Romano - Piazza Venezia - Campidoglio - Pantheon \\ 
   & Via del Corso - Isola Tiberina - Campo Marzio - Antico Ghetto \\ 
\hline
B3 & Aventino - Circo Massimo - Bocca della Verit\`a \\
   & Piramide Cestia - Porta San Paolo \\ \hline
C1 & Villa Albani - Via XX Settembre - Porta Pia\\
   & Galleria Borghese - Terme di Diocleziano \\ \hline
C2 & Stazione Termini - Santa Maria Maggiore - Via Nazionale \\
   & \conference 2012 venue - Via Cavour - San Pietro in Vincoli - Piazza Vittorio \\
   & Colosseo - San Clemente - Domus Aurea - Esquilino \\ \hline
C3 & San Giovanni - Terme di Caracalla - Porta Latina  \\ \hline
\end{tabular}
}
\caption{Lookup table used in the geometric approach of 
Section~\ref{se:divideetimpera}.  }
\label{tab:geometriclookup}
\end{figure*}

\noindent In spite of the restaurants above, you might still want to 
experiment with international cuisine:

\begin{enumerate}

\item[I\,1] {\em Hang Zhou}, Esquilino, Via Principe Eugenio, 82, 
ph: 06-4872732, Chinese (C2)

\item[I\,2] {\em Isola Puket}, Quartiere Africano, Via di Villa Chigi 
91, ph: 06-86212664, Thai (north of C1)

\item[I\,3] {\em Jaipur}, Trastevere, Via San Francesco a Ripa 56, 
ph: 06-5803992, the best Indian restaurant in town (A3)

\item[I\,4] {\em Thien Kim}, Centro Storico -- Campo De' Fiori, Via 
Giulia 201, ph: 06-68307832, Vietnamese (A2)

\end{enumerate}

\subsection{Corollaries}
\label{se:corollaries}

\vspace{-2mm}

After lunch or dinner, you can have a tasty gelato in one of the 
following celebrated ``gelaterie'':

\begin{enumerate}

\item[G1] {\em Il Gelato di San Crispino}, Centro Storico -- Tritone, 
Via Della Panetteria 42, ph: 06-6793924, closed on Tuesday (B1)

\item[G2] {\em Al Settimo Cielo}, Prati, Via Vodice 21a, ph: 
06-3725567, closed on Monday (A1)

06-3210807, closed on Monday (A1)

\item[G3] {\em Giolitti}, Via Degli Uffici del Vicario 40, ph: 
06-6991243 (B2)

\end{enumerate}

\noindent ... or you can drink a coffee at the following cafes (but 
please, remember, you are not supposed to order a cappuccino in Italy 
after noon!):

\begin{enumerate}

\item[C5] {\em Caff\`e Sant'Eustachio}, Centro Storico -- Pantheon, 
Piazza Sant'Eustachio 82, one of the best places to taste real 
Arabian Coffee. Try the caff\`e speciale, granita, irish cream, or 
parfait  (B2)

\item[C6] {\em Ciampini}, Centro Storico -- Campo Marzio, via della 
Fontanella di Borghese 59 / Piazza San Lorenzo in Lucina 29  (B2)
\end{enumerate}

\section{Divide et Impera (Divide and Conquer)}
\label{se:divideetimpera}

\vspace{-2mm}

Well, you know that Rome wasn't built in a day, and when the input 
size gets large, divide et impera can yield good results.
To apply the technique, divide the map of Rome into IX squares having 
approximately the same area, as shown in Figure~\ref{fi:mappa-large}. 
Based on your taste and preferences, choose an area where you would 
like to have dinner (e.g., area B2) and find the corresponding entry 
in the lookup table given in Figure~\ref{tab:geometriclookup}. Scan 
the list of restaurants of Section~\ref{se:hill-climbing}, discarding 
those not located in any of the areas listed in the selected lookup 
table entry. Choose a restaurant among the remaining ones: if the 
list of remaining restaurants is empty (don't panic: this is a low 
probability event), return ``Error \# MCCLXXIX: Diet suggested.''

\section{Random Sampling}
\label{se:sampling}

\vspace{-2mm}


If you choose a restaurant uniformly at random from the lists given 
in Section~\ref{se:localsearch} and in 
Section~\ref{se:hill-climbing}, you can still expect a reasonably 
good solution.

\section{The Greedy Approach}
\label{se:heuristic}

\vspace{-2mm}

If you want to try authentic Roman cuisine, look for the following 
dishes when you order (especially in traditional restaurants such as 
``trattorie'' and ``osterie''). A short guide can be probably helpful 
in order to get oriented:

\begin{description}

\item[Carciofi alla Giudia:] Roman-Jewish style artichokes, a must! 
You can have them in many Kosher restaurants in the Ghetto.

\item[Spaghetti cacio e pepe:] the real traditional roman dish:
pasta topped with grated pecorino Romano and ground black pepper.
It is also the favorite dish of Francesco Totti, the world-known
football player of A.S.~Roma, the football team of the
city.

\item[Abbacchio alla scottadito, abbacchio al forno:] ``burned 
finger'' lamb, roasted lamb.

\item[Filetti di baccal\`a:] deep fried cod.

\item[Fiori di zucca:] courgette flowers, typically stuffed with 
mozzarella and anchovies.

\item[Bucatini all'amatriciana:] pasta topped with tomato sauce, 
pancetta and a touch of black pepper.

\item[Bucatini alla gricia:] pasta topped with  
guanciale, pecorino cheese and a touch of black pepper. 
\item[Spaghetti alla carbonara:] pasta topped with egg, parmesan 
cheese, pancetta and black pepper.

\item[Saltimbocca alla romana:]  ``JumpInMouth'' Roman style, a thin 
fillet of veal topped with a slice of cured ham, white wine and sage.

\end{description}


\noindent And now, these are for the really adventurous people:

\vspace{-1mm}

\begin{description}

\item[Trippa alla romana:] tripe Roman style.

\vspace{-1mm}

\item[Rigatoni alla pajata:] pasta made with young veal calf 
intestines.

\vspace{-1mm}

\item[Coda alla vaccinara:] stewed oxtail.

\end{description}

\noindent Among other Italian foods that are worth trying, we also 
recommend: 

\begin{description}

\item[Appetizers:] bruschetta, crostini, suppl\`i, mozzarella di 
bufala, olive ascolane, antipasto all'italiana.

\item[Main courses:] tortellini or ravioli with meat (ringlets of 
dough filled with seasoned minced meat, typical from Bologna), trofie 
al pesto (a special recipe from Genoa), fettuccine\footnote{We refer 
the interested eater to Appendix I for details about the famous 
Fettuccine Alfredo.} alla boscaiola (fettuccine with mushrooms and 
peas), spaghetti alle vongole (pasta with clams).

\item[Desserts:] tiramis\`u (``PickMeUp,'' with coffee and mascarpone, 
an Italian soft cheese), cannoli or cassata siciliana (Sicilian 
pastry filled with cream and buttermilk curd), bab\`a napoletani 
(Neapolitan sponge-cake steeped in rum syrup), torta della nonna 
(grandma's cake, typically with cream and pine seeds).

\end{description}

\section{Experimental Evaluation of Pizzas in Rome}

\begin{figure}[t!]
\begin{center}
\includegraphics[width=0.4\textwidth]{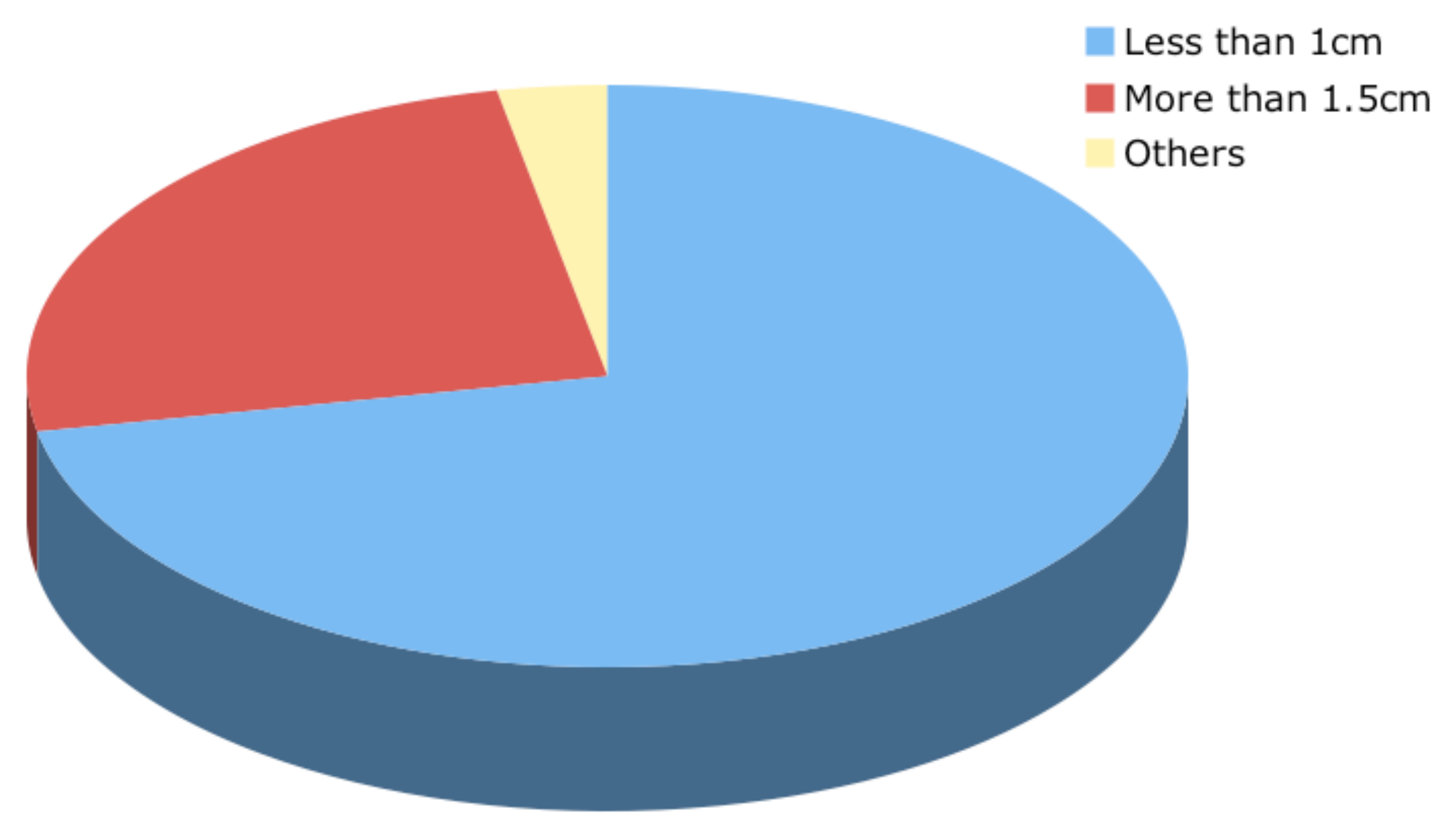}
\end{center}
\caption{Pizza Thickness Distribution.}
\label{fi:pizza}
\end{figure}

In Figure~\ref{fi:pizza} we report the results of our experiments on the
thickness of the pizzas we tried in Rome. We refer only to the pizzas
made in specialized restaurants, called \emph{pizzeria}, where the pizza
is served in dishes and it has the traditional round shape. On the other
hand there are many take-away shops called \emph{pizza a taglio}, where
the pizza is prepared in long, rectangular baking pans. It is important
to mention that the pizza in \emph{pizzerie} is mostly cooked in a
wood-fired oven, while in \emph{pizza a taglio} shops they usually cook
it in an electric oven.  As we can see from the plot, more than 70\% of
the pizzas\footnote{To be more precise with thickness of the pizza we
mean the thickness of the pizza base---that is, the pizza without the topping. } is less than one centimeter thick, and the remaining is much taller. It is interesting to notice that almost no intermediate thickness value has been found in our experiments. This confirms that in Italy there are two different kind of pizzas:
the \emph{romana (roman)} one and the \emph{napoletana (neapolitan)}. Here we summarize our main findings:
\begin{itemize}
\item The \emph{pizza romana}, besides being thinner, has a crispy base. On the opposite side, the \emph{pizza napoletana} has soft and pliable bases.
\item In Rome the pizza toped with tomato, mozzarella, anchovies, and oil is called Pizza Napoletana\footnote{The name Napoletana is obviously ambiguous: do we refer to the thickness or to the topping of the pizza? It is interesting to notice that, therefore, the interested eater can order a Pizza Napoletana Romana, meaning a thin pizza with the above topping, or a Pizza Napoletana Napoletana, always with the same topping but taller and softer.}. In Naples the same pizza is addressed as Pizza Romana\footnote{So in Naples it is possible to order a Pizza Romana Napoletana, but we strongly doubt it is possible to have a Pizza Romana Romana, because Neapolitans are really proud of their pizza!}. Surprisingly, there is no Pizza Romana in Rome and no Pizza Napoletana in Naples!

\item If you are from outside Italy and you like the ``pepperoni
pizza,'' don't ask for the ``pizza con i peperoni''---that is, pizza with
peppers! The proper name is ``Diavola.''
\end{itemize}

\section{Drinking Wine}

\begin{figure*}[t!]
\begin{center}
\includegraphics[width=0.65\textwidth]{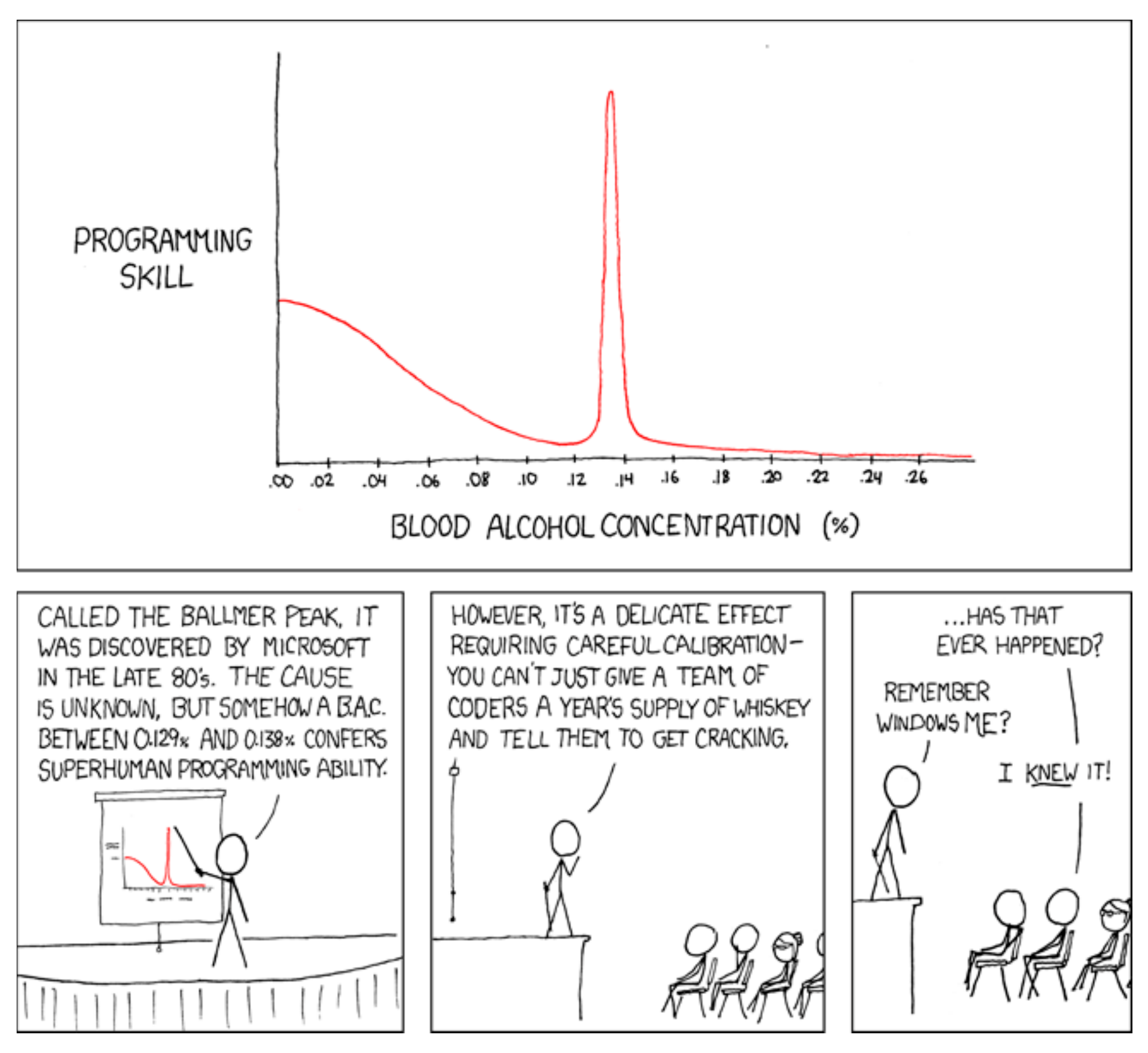}
\end{center}
\caption{A study on the relation between blood alcohol concentration and programming skills (from \texttt{xcdc.com}).\label{fi:xkcd}}\vspace{-4mm}
\end{figure*}

There is some experimental evidence that drinking alcoholic beverages during conference sessions can improve the appreciation of the talks (see also the related \emph{Ballmer Peak} shown in Figure~\ref{fi:xkcd}): results vary significantly whether the speaker, the audience or both are drinking (in a famous talk from FUN 2004 conference both the speakers and a large portion of the audience were drinking beer, that was distributed by the speakers (authors) that explained it by saying: ``\emph{we were drinking beer when we wrote this paper, and we think you should drink beer too when you listen to this talk}'').

We don't have enough space here to discuss this issue but, if you plan
to taste wine during \conference, we suggest the following list of
\emph{enoteche\footnote{The italian term \emph{enoteca} literaly means
  ``wine repository,'' and denotes a store in which you can buy wine; in many \emph{enoteche} it is also possible to taste wine like in a wine bar.}} and wine bars in Rome:
\begin{enumerate}
\item[W1] \emph{Trimani}, Via Cernaia, 37, ph: 06-4469630 (C1). This is probably the most famous roman \emph{enoteca}, whose roots date back to 1821. Here they only sell wine, but there is also \emph{Trimani Il Wine Bar}, Via Cernaia, 37, ph: 06-4469630, where you can taste wine and eat (C1)
\item[W2] \emph{Ai tre scalini}, via Panisperna, 251, ph: 06-48907495. Please note that this place, besides being there since 1895, it is in the street that gives his name to the famous \emph{Via Panisperna boys}, the group of young scientist led by Enrico Fermi, that in 1934 discovered the slow neutrons (B1)
\item[W3] \emph{Cavour 313}, Via Cavour, 313, ph: 06-6785496‎. It has a very large selection (C2)
\item[W4] \emph{Gusto}, Piazza Augusto Imperatore, 9, ph: 06-3226273. It has also a restaurant and a kitchen good store (A2)
\item[W5] \emph{Il pentagrappolo}, Via Celimontana, 21b, ph: 06-7096301. This is a wine \emph{music} bar (C3)
\item[W6] \emph{Cul de Sac}, Piazza di Pasquino, 73, 06-68801094. This is another very famous place in Rome (A2)
\end{enumerate}

If you prefer to drink \textbf{beer}, you can not miss the \emph{Beer~House}, Via Merulana 109 (it is very close to the \conference venue - R7 in Figure~\ref{fi:localsearch}). Here, besides many types of beers, you can enjoy the traditional home made roman cuisine that is well defined by the sign just above the door: \emph{cucina mi madre} that, in roman slang, means ``my mother is the chef'' (but we know that there is always something lost in translation). They have also free WiFi. 

We conclude by suggesting the nice (and very little!):
\begin{enumerate}
\item[W7] \emph{Caff\`e letterario Aquisgrana}, via Ariosto 28-30 
\end{enumerate}
that is very close to the \conference conference venue (few meters away): here you can taste not only wine, but also many beers and some traditional medieval alcoholic beverages: \emph{sidro, idromele and ippocrasso}! The last one, ippocrasso, is hand made by Simona and Patrizia, the owners of this place that offers also a small selection of books as well as handbound notebooks. And there is free WiFi!

\section{Eating SEAfood during \conference}

If you want to experiment SEAfood during the SEA conference, we provide a small list of restaurant below; however, if you also want to look at the SEA while eating SEAfood during the SEA conference, i.e., the \emph{full} experience, the recommended venue is Fiumicino, the small city on the coast near to Rome, close to the airport that bears its name. 

\begin{enumerate}
\item[S1] \emph{Baia Chia}, Via Machiavelli, 5, ph 06-70453452. Very close to the conference venue.
\item[S2] \emph{Mater Matuta},  Via Milano, 47, ph: 06-4823962. Also in walking distance from the conference venue.
\item[S3] \emph{Su Nuraghe}, Via Imperia, 66, ph: 06-44291846. It will serve you more food than you can eat!
\item[S4] \emph{Ragno d'Oro}, Via Silla, 26, ph: 06-3212362. Small, remember to book before you go!
\item[S5] \emph{Rosetta}, Rosetta, 8, ph: 06-6861002. Very very very expensive.  Don't say we didn't warn you!
\end{enumerate}

\section{Concluding Remarks and Open Problems}
\label{se: conclusions}

\begin{figure}[t!]
\begin{center}
\includegraphics[width=0.45\textwidth]{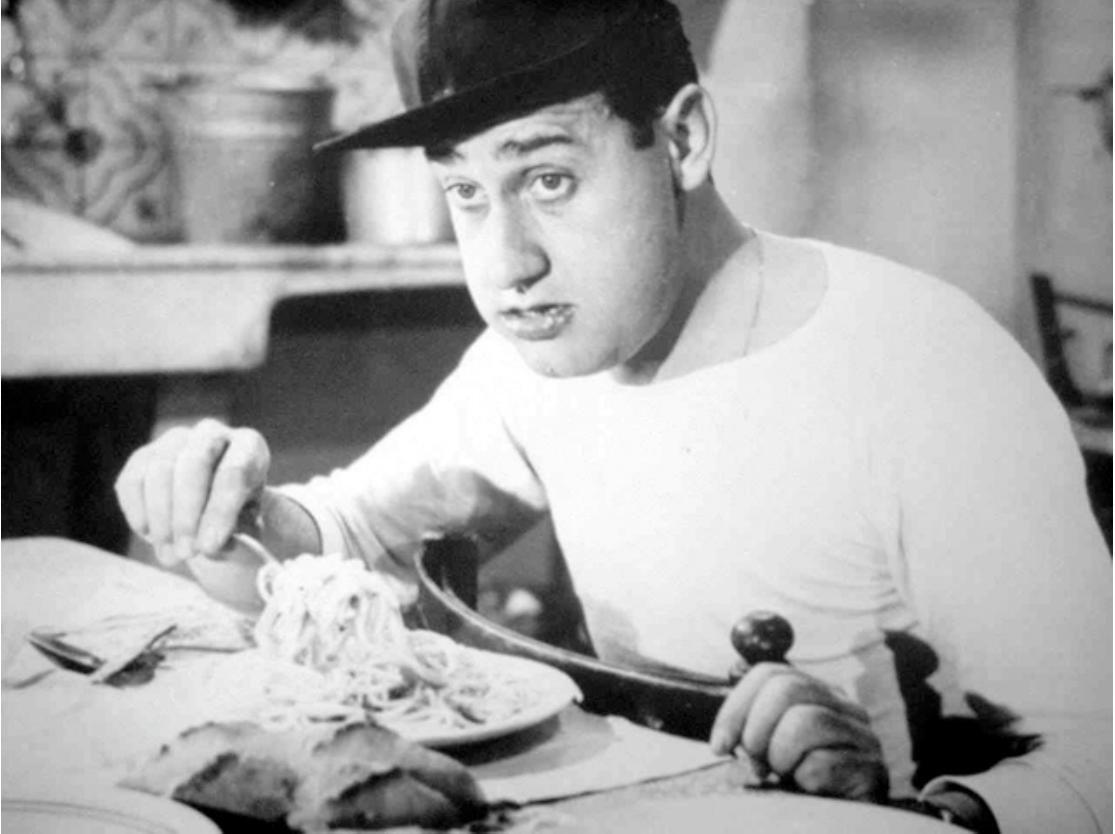}
\end{center}
\caption{A satisfied reader of this paper.}
\label{fi:albertone}
\end{figure}

We think that our experimental results should provide a good insight to
the problem addressed so far. However, more experiments are needed for
various related problems. For example, when a large group is at the end
of a nice dinner, and a cake is served, what is the right strategy? This
problem has been addressed in \cite{cake1,cake2}, but we strongly
believe that is missing an experimental validation of the theoretical results presented in the above papers. We hope that \conference participants will contribute in this direction. According to our experimental study done by the authors, 
the tested algorithms appear to achieve very good approximation ratios: the 
typical outcome of the experiments is exemplified in 
Figure~\ref{fi:albertone}. 


\noindent And to conclude, Horatius would say: \\

\vspace{2mm}

\centerline{{\em Nunc est bibendum,}}

\vspace{-2mm}

\begin{center}
{\em nunc pede libero pulsanda tellus}
\end{center}

\vspace{1mm}

\noindent ``Now it is the time to drink, now it is the time for the 
loose feet to hit the floor'' (Horatius, Carmina 1, 37).


\newpage
\section*{Appendix I}

\vspace{-2mm}


Wanna know the real story behind Fettuccine Alfredo?
In MCMXIV (1914) a small restaurateur living above his small Rome 
restaurant was
faced with a problem no respectable Roman could endure: the loss of 
his
wife's appetite. Many months pregnant, Alfredo di Lelio's wife 
refused all
her husband's best dishes. Watching his wife grow weaker, Alfredo 
made a
vow: ``I'll invent something so tasty that she will immediately 
succumb to
temptation,'' remembers Russell Bellanca, co-owner of the U.S. 
branches of
Alfredo the Original of Rome. So late one night in his tiny kitchen,
Alfredo dropped a handful of fresh fettuccine egg noodles into a pot 
of
boiling water. He melted butter and Parmegiano Reggiano cheese, and 
then
he mixed in the pasta. Fettuccine Alfredo was born! His wife cleaned 
her
plate, and a short time later, Alfredo II was born to the music of
customers downstairs in Alfredo's I's restaurant, all crying for his 
new
irresistible dish.
Today, of course, the original Alfredo's in Rome, which is in Piazza 
Augusto Imperatore, is lorded over by Alfredo
the III. Amazingly enough, nowadays fettuccine Alfredo are more known 
in the U.S. than in Rome: there are branches of Alfredo's in 
Manhattan and Miami Beach, as
well as in Disney World's Epcot Center.

\end{document}